\documentclass[preprint2, times]{aastex6}
\pdfoutput=1

\usepackage{natbib}
\usepackage{amsmath}
\usepackage{amssymb}
\usepackage{subfigure}
\usepackage{url}
\usepackage{lineno}

\renewcommand{\vec}[1]{ {\mathbf #1} }

\newcommand{\grad}{ {\bf \nabla } }
\newcommand{\curl}{ {\bf \nabla} \times}

\newcommand{\Eq}{{Equation}}

\newcommand{\Fig}{{Figure}}

\newcommand{\dive}{\nabla\cdot}
\newcommand{\divB}{\nabla\cdot\vec{B}}

\providecommand{\dodoi}[1]{doi:~\href{http://doi.org/#1}{\nolinkurl{#1}}}
\providecommand{\url}[1]{\href{#1}{#1}}
\providecommand{\doeprint}[1]{\href{http://ascl.net/#1}{\nolinkurl{http://ascl.net/#1}}}
\providecommand{\doarXiv}[1]{\href{https://arxiv.org/abs/#1}{\nolinkurl{https://arxiv.org/abs/#1}}}

\graphicspath{{}}

\begin{document}


\title{Numerical Simulation of Solar Magnetic Flux Emergence Using the
  AMR--CESE--MHD Code}

\author{ Zhipeng Liu\altaffilmark{1}, Chaowei Jiang\altaffilmark{1}, Xueshang Feng\altaffilmark{1},  Pingbing Zuo\altaffilmark{1}, Yi Wang\altaffilmark{1}}

\altaffiltext{1}{Institute of Space Science and Applied Technology,
  Harbin Institute of Technology, Shenzhen 518055, China; chaowei@hit.edu.cn
}

\begin{abstract}
Magnetic flux emergence from the solar interior to the atmosphere is believed to be a key process of formation of solar active regions and driving solar eruptions. Due to the limited capability of observation, the flux emergence process is commonly studied using numerical simulations. In this paper, we developed a numerical model to simulate the emergence of a twisted magnetic flux tube from the convection zone to the corona using the AMR--CESE--MHD code, which is based on the conservation-element solution-element method with adaptive mesh refinement. The result of our simulation agrees with that of many previous ones with similar initial conditions but using different numerical codes. In the early stage, the flux tube rises from the convection zone as driven by the magnetic buoyancy until it reaches close to the photosphere. The emergence is decelerated there and with piling-up of the magnetic flux, the magnetic buoyancy instability is triggered, which allows the magnetic field to partially enter into the atmosphere. Meanwhile, two gradually separated polarity concentration zones appear in the photospheric layer, transporting the magnetic field and energy into the atmosphere through their vortical and shearing motions. Correspondingly, the coronal magnetic field has also been reshaped to a sigmoid configuration containing a thin current layer, which resembles the typical pre-eruptive magnetic configuration of an active region. Such a numerical framework of magnetic flux emergence as established will be applied in future investigations of how solar eruptions are initiated in flux emergence active regions.

\end{abstract}


\section{Introduction}
\label{sec:intro}

Coronal mass ejections (CMEs), flares and jets are the major forms of eruptions in solar activities, and the physical mechanisms of their trigger and driver are an important research topic in solar physics. Numerous observational studies have reported that these eruptive activities frequently occur in solar active regions, and it is generally believed that the core structure of the pre-eruptive field is in the form of either a twisted flux tube, i.e., a magnetic flux rope (MFR) or a strongly sheared magnetic arcade \citep{Green2011PhotosphericFC, Patsourakos2013}. The entire pre-eruption configuration consists of the core field (either an MFR or a sheared arcade) and an envelope field (overlying field) that confines the core field, while eruptions occur when some kind of instabilities destabilize their force balance \citep{Archontis2012}.

It is currently accepted that solar active regions are formed by magnetic flux emergence, the process of magnetic fields generated by solar dynamo entering the solar atmosphere from the depths of the convection zone, which is also considered to be one of the key mechanisms in producing solar eruptive activity \citep{DG2015EAR, ChenP2011}. Although the emerging magnetic field has been thought to be sufficient in itself to generate an eruption \citep{Dmoulin2002WhatIT, Nindos2003}, in many cases it acts as a trigger for a pre-existing eruptive configuration \citep{Feynman1995TheIO, Williams2005EruptionOA}. In a stable pre-eruption configuration, the upward magnetic pressure of the internal flux rope is in equilibrium with the downward tension of the envelope field \citep{Archontis2012, Leake2013}. When a new flux emerges in the vicinity of the pre-existing eruption configuration, their interaction causes magnetic reconnection that could reduce the tension of the envelope field and lead to the eruption \citep{Chen2000}. There are two possible ways of reconnection operating in this process, which are tether-cutting \citep{Moore1992} and breakout reconnection \citep{Antiochos1999}.
In other cases, the pre-eruption configuration is associated with the ideal instability. Continuous flux emergence may push the magnetic configuration higher, and when the envelope field decays too fast with height, the MFR will run into the torus instability and erupt \citep{Kliem2006}. Flux emergence can also increases the degree of twist of the MFR, and when a certain value is exceeded, it triggers kink instability and an eruption \citep{Anzer1967, Torok2004}.

Since without a direct observational probe of the dynamics of magnetic flux emergence from below the solar surface (i.e., the photosphere), many efforts have been devoted to numerical magnetohydrodynamic (MHD) simulations of the flux emergence. As pioneered by the early work of Shibata and colleagues \citep{Shibata1989}, a large number of works of flux emergence simulation (FES) have been carried out, in particular, for mimicking simulations of a twisted flux tube emergence into the solar atmosphere \citep{ Morenoinsertis1996TheRO, Fan2001, Magara2001a, Arber2001a, Manchester2004Emergence, Archontis2004, Murray2006, Leake2006, Toriumi2010, Cheung2014, Syntelis2017, Toriumi2019, Fan2021}. These simulations have successfully reproduced some of the observed phenomena, such as the vortical motion of the emerging polarities on the photosphere, the sigmoid shaped coronal MFR, and these comparative results confirm the reliability of MHD simulations. 

To simulate a flux tube emerging from the convection zone into the corona (and to further study how it erupts) requires the numerical model to incorporate the highly stratified solar plasma including all the different layers from below the surface, to the photosphere, the chromosphere, the transition region, and the corona, which have physical behaviors rather different from each other. Therefore, a major challenge in self-consistent simulations of magnetic flux emergence to its eruption is to resolve the multiple spatial and temporal scales in a single model. For example, near the photosphere, the scale heights (of gas pressure) are only about one hundred kilometer, and the gas density varies by more than eight orders of magnitude within a few megameters, while in the corona the scale height is tens of megameters, i.e., with nearly three orders of magnitude larger than the photospheric ones. On the contrary, the time scales of evolutions in the photosphere and below, in which the magnetic field is controlled mainly by the plasma, are much longer than that in the corona, in which the plasma is controlled by the magnetic field. Thus, most of the current 3D FESs chose to use relatively small computational domains of a few tens of megameters in three spatial directions and short time durations of, typically, a few hours. The burden of computational resources would be too much if one wants to simulate the long-term (e.g., days) evolution of active region size (e.g., hundreds of megameters). 

The motivation of this paper is to develop a new numerical model of magnetic flux emergence by using our AMR--CESE--MHD code~\citep{Jiang2010}, in particular, utilizing the features of adaptive mesh refinement \citep[AMR,][]{BERGER198964}. The technique of AMR has been developed rapidly in computational fluid dynamics and is becoming a standard tool for treating problems with multi-orders of spatial or temporal scales, which fits well for FES. By automatically adapting the computational mesh to the solution of the governing partial differential equations (PDEs), methods based on AMR can assign more mesh points for regions demanding high resolution (e.g., high gradient regions) and at the same time, give fewer mesh points to other less interested regions (low gradient regions), thereby providing the required spatial resolution while minimizing memory requirements and CPU time. Although many classical numerical MHD solvers based on either finite difference or finite volume methods have been used in previous FESs, such as ZEUS--3D code \citep{Stone2008}, the modified Lax-Wendroff method \citep{Magara2001a, Toriumi2010}, and the Lagrangian remap scheme \citep[Lare3d,][]{Arber2001a}, few of these FESs have implemented with the AMR. There are only two simulations used AMR \citep{Cheung2006, Martinez-Sykora2015}, but both these two early simulations only studied the evolution of the flux tube below the photosphere and in \cite{Cheung2006} the simulation is carried out within 2.5D rather than 3D. On the other hand, the CESE method is distinct from the classical numerical methods of the finite-difference or finite-volume schemes, as it has a much simplicity in mathematics without Riemann solver or eigen-decomposition, but can achieve higher accuracy at equivalent grid points, which is also desirable for the FES. The AMR--CESE--MHD code has achieved many excellent results in other simulations, such as in analysis of the fundamental initiation mechanism of solar eruptions \citep{Jiang2021,Bian2022}, the data-driven active region evolution and eruptions \citep{Jiang2016NC, Jiang2021b, Jiang2022DatadrivenMO}, and the solar wind modellings \citep{Feng2012SoPh}.

In this paper, we report our first step of implementation of applying the AMR--CESE--MHD code to FES, by simulating the emergence of a twisted flux tube in a simply stratified solar atmosphere from the convection zone to the corona. In the following, Section \ref{sec:method} describes the details of the model and numerical methods. In Section \ref{sec:ressult}, we show the process and key features of the 3D magnetic flux emergence, which is overall consistent with previous FESs. In Section \ref{sec:sum}, we summarize and give outlooks for future study based on the new FES model.

\section{Model}
\label{sec:method}

\subsection{Initial conditions}
\label{sec:init}
The initial settings of our model are similar to that used in typical
simulations of the emergence of twisted flux tube from below the
photosphere to the corona, and particularly the parameters are mostly
close to the values used in \citet{Fan2009}. The simulation volume is
a Cartesian box of $-14.4$~Mm $\le x \le 14.4$~Mm, $-14.4$~Mm
$\le y \le 14.4$~Mm and $0 \le z \le 28.8$~Mm where the $z$ axis
is the height with $z=0$ denoting the lower boundary, which is a depth of $4.5$~Mm below the photosphere.

The initial conditions consist of a plasma in hydrostatic equilibrium stratified
by solar gravity with a characteristic temperature profile from the
top layer of convection zone to the corona, which is given by a
piece-wise function of height

\begin{equation}\label{eq:T}
  T(z) = \left\{ \begin{array} {r@{\quad:\quad}l}
      T_{\rm ph} -\frac{\gamma -1}{\gamma}\frac{g_0}{R} (z-z_{\rm ph}) & z \le z_{\rm ph} \\
      T_{\rm ph} & z_{\rm ph} < z \le z_{\rm ch} \\
      T_{\rm ph}\left( \frac{T_{\rm cor}}{T_{\rm ph}}
      \right)^{\frac{z-z_{\rm ch}}{z_{\rm cor}-z_{\rm ch}}}  & z_{\rm ch} <z \le z_{\rm cor}\\
      T_{\rm cor} & z \ge z_{\rm cor}
      \end{array}\right. ,
\end{equation}
where the photospheric temperature is $T_{\rm ph} = 5\times 10^{3}$~K
and the coronal temperature $T_{\rm cor}=10^{6}$~K. The heights
$z_{\rm ph} = 4.5$~Mm, $z_{\rm ch} = z_{\rm ph} + 1.5$~Mm, and
$z_{\rm cor} = z_{\rm ch} + 1.5$~Mm are the heights of photosphere,
chromosphere, and the base of the corona, respectively. As the
modelling region is a small range around the solar surface, we assume
that the solar gravity is constant with value of
$g_0 = 274$~m~s$^{-2}$ on the solar surface, and
$R = 8.25\times 10^{3}$ is the gas constant such that $p = \rho R T$.
By assuming a balance of pressure gradient and gravity,

\begin{equation}\label{eq:static}
  \frac{d p}{d z} = -\rho g = -\frac{p}{R T/g},
\end{equation}
the pressure is
 
\begin{equation}\label{eq:T}
  p(z) = \left\{ \begin{array} {l}
      p_{\rm ph}\left[\frac{T(z)}{T_{\rm
                   ph}}\right]^{-\frac{g_{0}}{k_{\rm c}R}}, z \le z_{\rm ph} \\
      p_{\rm ph}\exp \left[\frac{g_{0}(z_{\rm ph}-z)}{RT_{\rm ph}}\right], z_{\rm ph} < z \le z_{\rm ch} \\
      p_{\rm cor} \exp\left[ \frac{z_{\rm cor}-z_{\rm ch}}{RT_{\rm
        ph}/g_{0}\ln(T_{\rm cor}/T_{\rm ph})}
    \left(\frac{T_{\rm ph}}{T(z)} -\frac{T_{\rm
        ph}}{T_{\rm cor}}\right) \right],  \\ z_{\rm ch} <z \le z_{\rm cor}\\
      p_{\rm cor}\exp \left[\frac{g_{0}(z_{\rm cor}-z)}{RT_{\rm cor}}\right], z \ge z_{\rm cor}
      \end{array} \right. ,
\end{equation}
where

\begin{eqnarray}
p_{\rm ph} = p_{\rm cor}\exp\left[ \frac{z_{\rm cor}-z_{\rm ch}}{RT_{\rm
        ph}/g_{0}\ln(T_{\rm cor}/T_{\rm ph})}
    \left(1 -\frac{T_{\rm
        ph}}{T_{\rm cor}}\right) \right] \notag \\ 
\exp\left[\frac{g_{0}(z_{\rm ch}-z_{\rm ph})}{RT_{\rm ph}}\right].
\end{eqnarray}
Then we place a uniformly twisted magnetic flux tube below the
photosphere. It is oriented along the $x$-direction with a straight
axis located at $y = 0$ and $z = 2.4$~Mm (i.e., $2.1$~Mm below the
photosphere surface). In the local cylindrical coordinate system
centered at the tube axis, its magnetic field is given by

\begin{equation}\label{fluxtube}
  \vec B = B_x (r) \hat{\vec x} + B_\theta (r) \hat{\mathbf{\theta}},
\end{equation}
where

\begin{equation}\label{fluxtubeBx}
  B_x (r) = B_0 \exp (-r^2/a^2), \ \
  B_\theta(r) = q r B_x(r).
\end{equation}

\begin{figure}[htbp]
  \center
  \includegraphics[width=8cm]{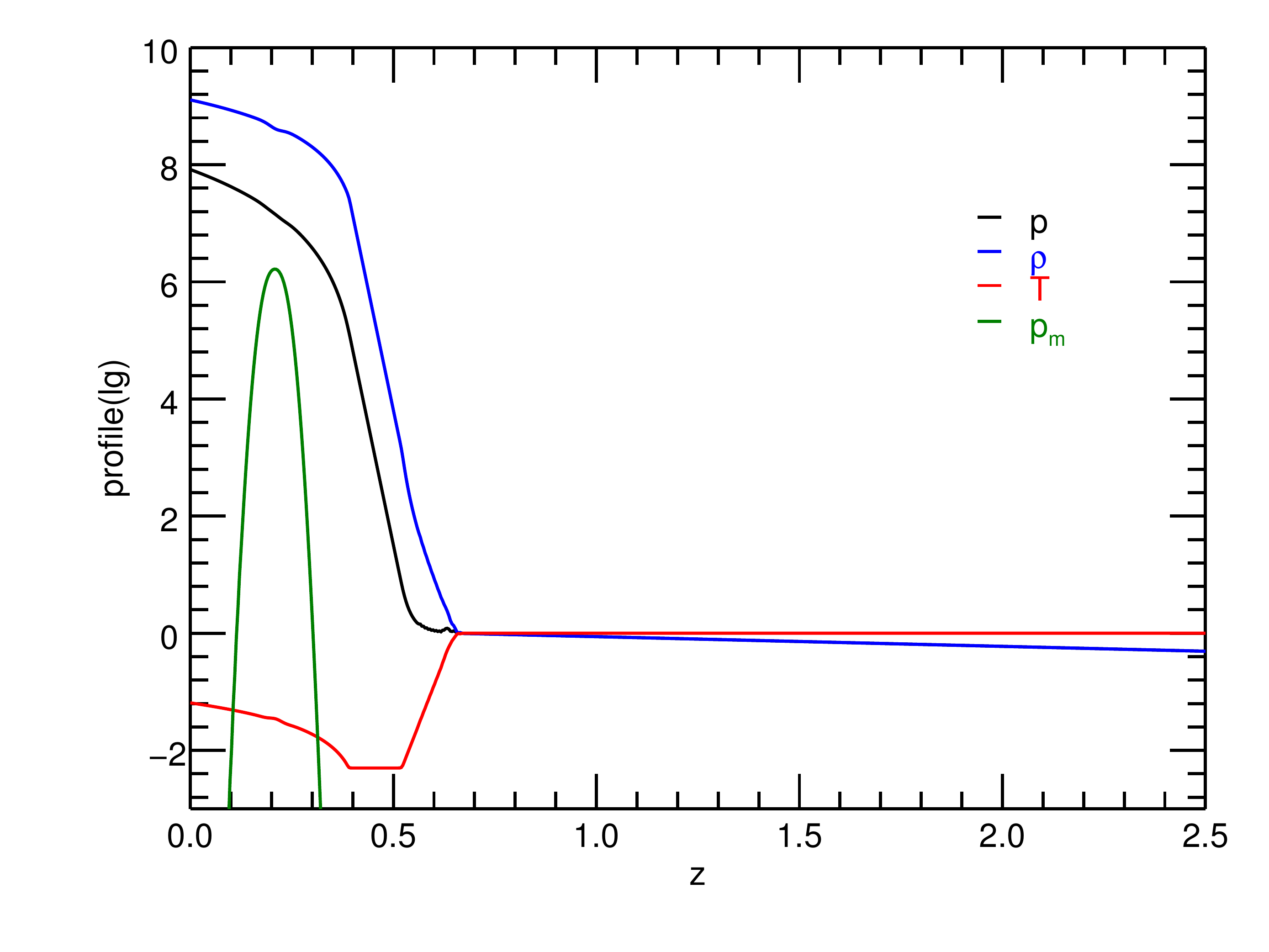}
  \caption{The vertical profiles of plasma pressure, density, and temperature, and the magnetic pressure through the central vertical line $(x, y) = (0, 0)$.}
  \label{Fig:profile}
\end{figure}

\begin{figure}[htbp]
  \center
  \includegraphics[width=8.5cm]{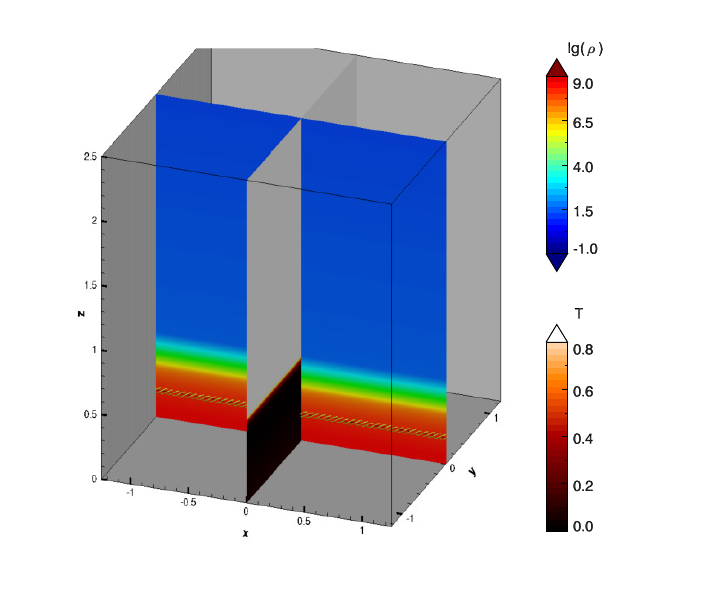}
  \caption{The simulation domain, with different slices to denote
    the distribution of density ($x = 0$) and temperature ($y = 0$). The flux tube is
    shown by the field lines. }
  \label{Fig:domain}
\end{figure}
In the above equations, $\hat{\vec x}$ is the tube axial direction,
$\hat{\mathbf{\theta}}$ is azimuthal direction in the tube cross
section, and $r$ denotes the radial distance to the tube axis. For the specific values of the parameters, we set $B_{0} = 3400$~G, and the radius $a=375$~km. The twist parameter is set as $q=-1/a$ which is the threshold value for the kink instability \citep{Linton1996}. The plasma $\beta$ (defined as the ratio of the plasma pressure to the magnetic pressure) is $8.99$ at the axis of the tube.

In \Fig~\ref{Fig:profile} we show the vertical profiles of plasma pressure, density, temperature, and the magnetic pressure through the central vertical line $(x, y) = (0, 0)$. Note that all the parameters in the figures of this paper are divided by their values in the coronal base ($z = z_{\rm cor}$).

Since the flux tube is not force-free and to make it in equilibrium, 
we modified the plasma pressure  (without changing the density) in the
tube by a difference of

\begin{equation}\label{p1}
  p_1(r) = \frac{B_{x}^{2}(r)}{2} \left[q^{2}(\frac{a^{2}}{2} - r^{2})-1\right],
\end{equation}
which ensures that the Lorentz force is balanced by the gas pressure
gradient. Then to make the flux tube buoyant, we further added a density
change $\rho_{1}$ within the tube

\begin{equation}\label{rho1}
  \rho_1 = -\rho_{0}(z) \frac{B_{x}^{2}(r)}{2p_{0}(z)}
  \left[(1+\epsilon)\exp(-x^{2}/\lambda^{2})-\epsilon\right],
\end{equation}
where $\rho_{0}$ is the background density, $\epsilon = 0.2$ and
$\lambda=1.2$~Mm. This makes the middle portion of the flux tube
buoyant since the modified density is lower than the background one
(see \Fig~\ref{Fig:domain}). Note that the buoyancy declines with horizontal
distance from $x = 0$ following a Gaussian profile with an $e$-folding
length of $\lambda$, and the two ends of the tube are slightly anti-buoyant. This can let the tube to emergence more vertically when it crosses the photosphere.

\subsection{MHD equations}
\label{sec:model}

We numerically solve the full set of MHD equations with the above
initial conditions. Before describing the model equations in the code,
it is necessary to specify the quantities used for
non-dimensionalization. Different from many other papers of FES, here we use typical values in the coronal base ($z=z_{\rm cor}$) for normalization, since our future application of this model will be devoted mainly to investigation of the eruptions in the corona as driven by the flux emergence. The specific values for all the variables and parameters are listed in Table~\ref{table1}:

\begin{table}[htbp]
  \centering
  \begin{tabular}{lll}
    \hline \hline
    Variable &  Expression & Value \\
    \hline
    Density        & $\rho_{s}$ =  $n m $ & $3.35 \times 10^{-15}$~g~cm$^{-3}$ \\
    Temperature    & $T_{s}$       & $1\times 10^{6}$~K \\
    Length         & $L_{s}$  = $16$~arcsec  &  $11.52$~Mm\\
    Pressure       & $p_{s}=2nk_{B}T_{s}$   &  $2.76\times 10^{-2}$~Pa \\
    Magnetic field & $B_{s}=\sqrt{\mu_{0}p_{s}}$ & $1.86$~G\\
    Velocity       & $v_{s}=\sqrt{p_s / \rho_{s} }$ & $90.9$~km~s$^{-1}$ \\
    Time           & $t_{s}=L_{s}/v_{s}$ & $127$~s \\
    Gravity        & $g_{s}=v_{s}/t_{s}$ & $1.05$~km~s$^{-2}$\\
    \hline
  \end{tabular}
  \caption{Parameters used for non-dimensionalization. $n$ is a
    typical value of electron number density in the corona given by
    $n=1\times 10^{9}$~cm$^{-3}$ and $m$ is the mean atomic mass,
    which is 2 times of proton mass.}
  \label{table1}
\end{table}

In the rest of the paper all the variables and quantities are written
in non-dimensionalized form if not specified. As such, the full set of
MHD equations are given as

\begin{align}
  \label{eq:MHD}
  \frac{\partial \rho}{\partial t}&= -\dive (\rho\vec v) ,
  \nonumber \\
  \rho\frac{D\mathbf{v}}{D t} &= -\grad p+\vec J\times \vec B+\rho\vec
  g + \nabla\cdot(\nu\rho\nabla\mathbf{v}) - \vec B \divB,
  \nonumber \\
  \frac{\partial \vec B}{\partial t} &=
  \curl (\vec v \times \vec B) + \grad (-\mu \divB) - \vec v \divB,
  \nonumber \\
  \frac{D T}{D t} &=
  (1-\gamma)T\nabla\cdot\vec v - \nu_{T}(T-T_{0}).
\end{align}
where $\vec J = \nabla \times \vec B$ is the current density, and
$\gamma = 5/3$ is the adiabatic index. Note that we artificially add a source term
$-\nu_{T}(T-T_0)$ to the equation of temperature, where $T_0$ is the
temperature at the initial time $t=0$, and $\nu_{T}$ is a prescribed
coefficient given by

\begin{equation}
  \label{eq:Tsource}
  \nu_{T} = \frac{1}{2}\left[1- \tanh\left(\frac{T-T_{\min}}{T_{\min}}\right) \right],
\end{equation}
where $T_{\rm min} = T_{\rm ph}/2 = 2.5\times 10^{-3}$. This source
term is a Newton relaxation of the temperature to its initial value by
a time of $1/\nu_{T}$, and with the specific choice of $\nu_{T}$, it
is aimed to avoid over cooling of the plasma during the fast expansion
of flux tube after it passes through the photosphere into the corona. As our code
has rather small numerical diffusion, we need some additional kinetic
viscosity to dissipate the small-scale disturbances arisen in the simulation. We use a small viscosity
coefficient $\nu = 0.1 \Delta x v_{\max} $, which is given according
to the local grid size $\Delta x$ and the local largest wave
speed $v_{\max}$

\begin{equation}
  \label{eq:vmax}
  v_{\max} = v + \sqrt{c_{\rm s}^{2} + v_{\rm A}^{2}},
\end{equation}
where $v$, $c_{\rm s} = \sqrt{\gamma p/\rho}$, and
$v_{\rm A} = B/\sqrt{\rho}$ are the motion speed, the sound speed, and
the Alfv{\'e}n speed, respectively. It corresponds to a grid Reynolds number of

\begin{equation}
  \label{eq:Rg}
  R_{\rm g} = \frac{\Delta x^{2} / \nu }{\Delta x/v_{\max} } = 10.
\end{equation}

In the MHD equation, all the terms associated with $\divB$ are employed to eliminate the
$\divB$, or the magnetic monopole, which should be exactly zero but
arises due to numerical errors. The diffusion coefficient for magnetic
monopole is given by $\mu = 0.4(\Delta x)^{2}/\Delta t$ according to
the local grid size and time step. Finally, we note that in the
magnetic induction equation no explicit resistivity is used, but
magnetic reconnection is still allowed through numerical diffusion
when a current layer is sufficiently narrow such that its thickness is
close to the grid resolution \citep[see also][]{Jiang2021}.

\subsection{Numerical scheme and grid setting}
\label{sec:num}

\begin{figure}[htbp]
  \center
  \includegraphics[width=8cm]{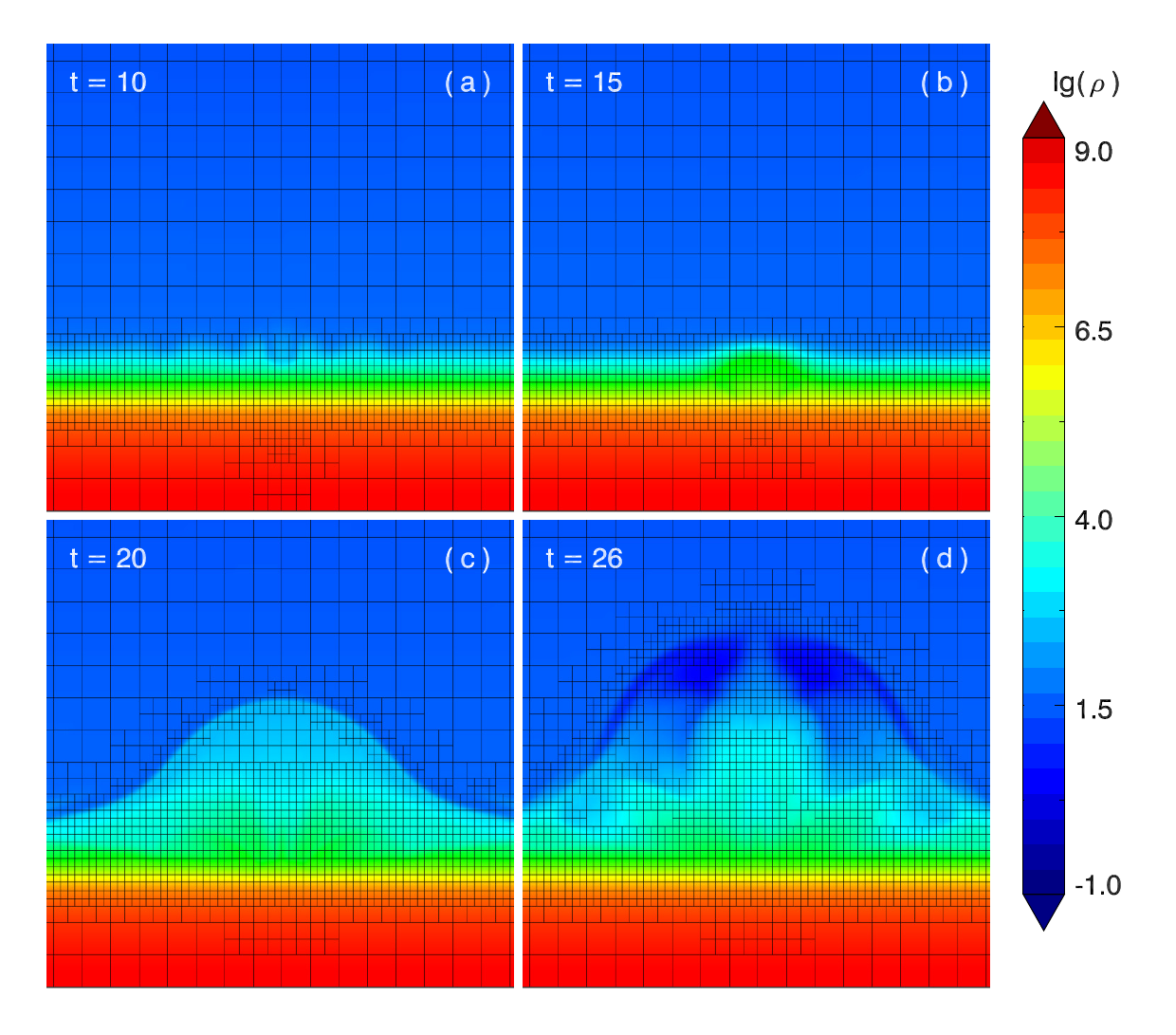}
  \caption{The evolution of the block (which contains $ 8^3 $ cells) distribution during the simulation. The whole computation domain is divided into blocks of different sizes, which are shown by the grid lines. Each block further  consists of $ 8^3 $ cells (i.e., 8 in each direction), which is not shown in the figure because it will be too dense to show. The color indicates the distribution of the logarithm of the density $\rho$ on slice $x = 0$.}
  \label{Fig:AMR}
\end{figure}

The above full set of MHD equation~(\ref{eq:MHD}) is solved by the CESE--MHD code~\citep{Jiang2010}. The CESE method treats time as being a dimension similar to the three dimensions in space when solving the 3D time-dependent governing equations. By reasonably introducing the conservation element (CE) and the solution element (SE) in the 4D space of space and time, and using the conservation law to compute the space-time flux to obtain the information of the next time. Contrast with many other numerical schemes, the CESE method is simple in mathematics since it does not need Riemann solver or eigen-decomposition, but can achieve higher accuracy with the equal number of grid points. More details of the scheme can be found in \citet{Feng2007, Feng2006} and \citet{Jiang2010, Jiang2011s}.



The simulation volume is resolved by an AMR
grid of cubic cells. The AMR is designed to automatically and
dynamically resolve with highest resolution the large gradient regions
in plasma variables arisen during the magnetic flux emergence as well
as the strong magnetic field regions, mainly in the flux
tube. Specifically, the whole volume is divided into blocks of
different sizes, and each block consists of $8^{3}$ cells, and after
each time step of advancing solution, we checked whether the blocks
need to be refined or coarsened by a set of physics-based criteria,
which are defined as

\begin{eqnarray}
  \label{eq:amr_crit}
  \chi_{\rho} = \Delta_{\rm c} \frac{ |\nabla \rho|}{\rho}, \ \
  \chi_{p} = \Delta_{\rm c} \frac{ |\nabla p|}{p}, \nonumber \\
  \chi_{B} = \Delta_{\rm c} \frac{ |\nabla (p_{B})|}{p_{B}}, \ \
  \chi_{T} = \Delta_{\rm c} \frac{ |(\vec B\cdot\nabla)\vec B|}{p_{B}},
\end{eqnarray}
where $p_{B} = B^{2}/2$ is the magnetic pressure, and $\Delta_{\rm c}$ is the length of the cell. If any of the four quantities in any cell of a block is larger than the threshold for refinement, which are given as $0.25$ for both $\chi_{\rho}$ and $\chi_{p}$ and $0.1$ for both $\chi_{B}$ and $\chi_{T}$, this block will be refined. On the other hand, if all the four quantities in all cells of a block are smaller than the threshold of coarsening, which are $0.1$ for both $\chi_{\rho}$ and $\chi_{p}$ and $0.04$ for both $\chi_{B}$ and $\chi_{T}$, this block will be coarsened. After the refinement and coarsening, the variables on the new grid will be interpolated from the old grid and then the solution will be advanced in the new time step. Note that the application of the two criteria $ \chi_{B}$ and $\chi_{T}$, which is associated with magnetic field $\vec B$ only, is restricted within the strong-field region satisfying $p_{B}/\rho > 1\times 10^{-3}$ and $p_{B}/p > 1\times 10^{-3}$ . \Fig~\ref{Fig:AMR} shows the evolution of the block (which contains $ 8^3 $ cells) distribution with these criteria applied during the simulation. We use three levels of AMR with the highest resolution of $45$~km and lowest resolution of $180$~km. Then we employed the PARAMESH software package~\citep{MacNeice2000} to manage the AMR procedure and the paralleling computing.

Since the spatial resolutions and the wave speeds of blocks within the
computational domain vary significantly, the timesteps computed using
a fixed Courant number $ C \sim 1$,
$\Delta t = C \Delta_{\rm c}/w$, where $w$ is the maximal wave
speed in the block, will also vary significantly. A simple way is to
use a uniform timestep for all the blocks, which is defined as
$\Delta t_{\rm g} =  C \Delta_{\min}/w_{\max}$ where
$\Delta_{\min}$ is the highest resolution and $w_{\max}$ is the maximal
wave speed in the entire computation domain. However, this will
increase significantly the numerical diffusion on the coarser blocks
and in the low wave speed areas, especially evident contrasting the
wave speeds (mainly the sound speed) in the photosphere and in the
corona, since the local timestep $\Delta t$ is much larger than the
global one $\Delta t_{\rm g}$, or in other words, the local Courant
number defined as
${C}_{\rm l} = w\Delta t_{\rm g}/\Delta_{\rm c} $ is much smaller than unity. This problem is especially serious for the CESE scheme which is
sensitive to the local Courant number. To overcome this problem, we
use time marching with block-based variable timestep, in which
different timesteps are used for different blocks, with the timesteps
defined as $\Delta t = {C} \Delta_{\rm c}/w_{\max}$ thus directly
proportional to the resolutions of the blocks. Furthermore, we use the
Courant number insensitive (CNIS) approach~\citep{Chang2005} which can reduce
the numerical dissipation substantially in the case that the local
Courant number is small.

\section{Result}
\label{sec:ressult}

\subsection{General evolution}
\label{sec:overview}

\begin{figure*}[htbp]
  \center
  \includegraphics[width=18cm]{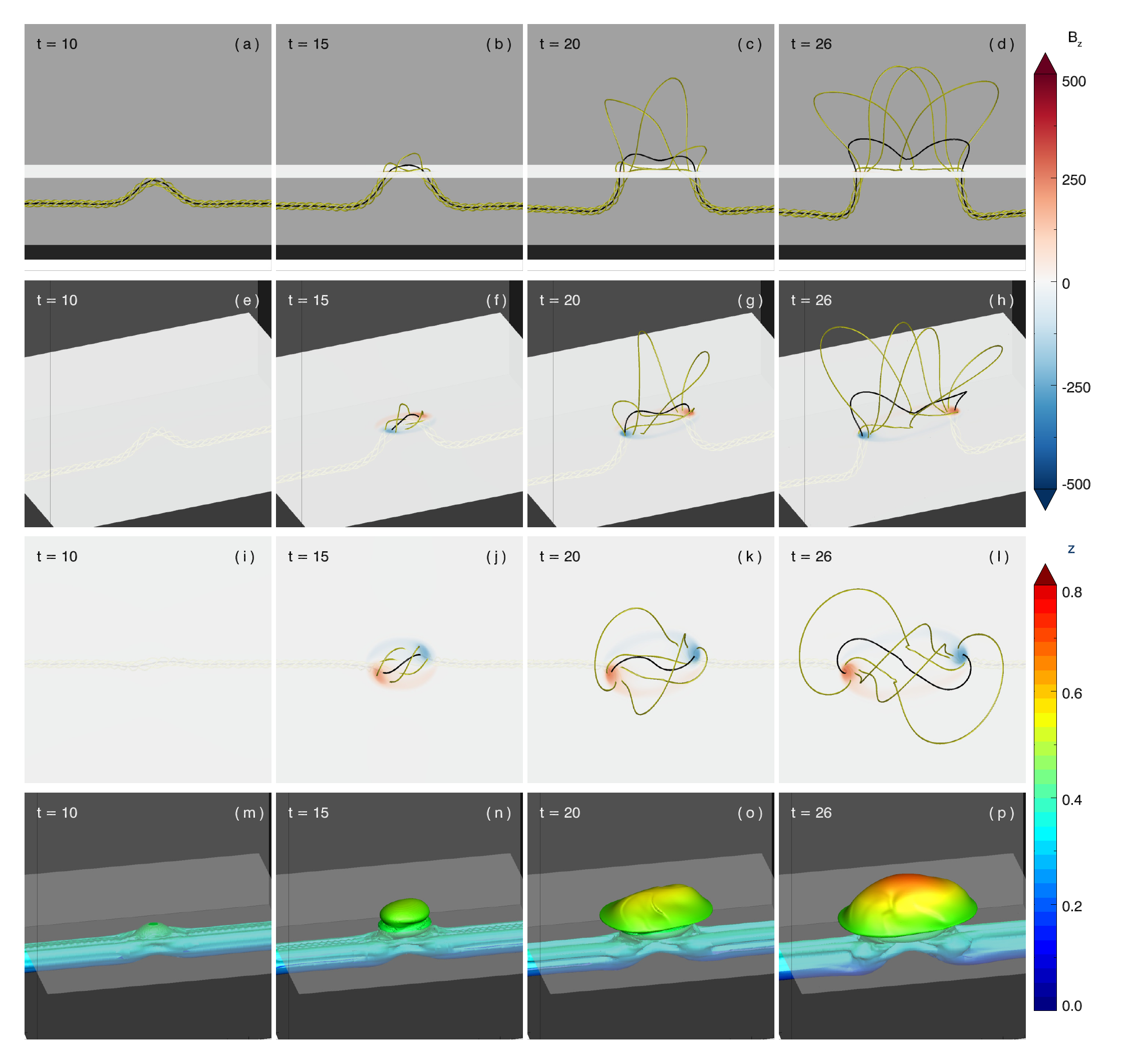}
  \caption{(a-l)  Three perspective views of 3D structure and evolution of the magnetic flux tube during the emergence. (m-p) The iso-surface of the flux tube with magnetic field strength $B = 0.1B_0$. }
  \label{Fig:Bline}
\end{figure*}

The whole process of subsurface twisted magnetic flux tube emerging in the atmosphere is consistent with previous simulations \citep{Fan2001, Fan2009, Manchester2004Emergence, Archontis2004, Magara2004, Murray2006, Leake2006, Leake2013, Syntelis2017}. The middle section of the flux tube starts to rise upward from the convection zone due to the magnetic buoyancy as caused by density deficit, while the two ends of the tube sink slightly because of artificial anti-buoyancy. The middle part of the tube continues to rise and expand with height until its apex touches the surface. Then the accumulation of the magnetic field under the surface triggers the magnetic buoyancy instability, allowing part of the flux to enter the photosphere/chromosphere and expand rapidly in the corona.

\begin{figure*}[htbp]
  \center
  \includegraphics[width=18cm]{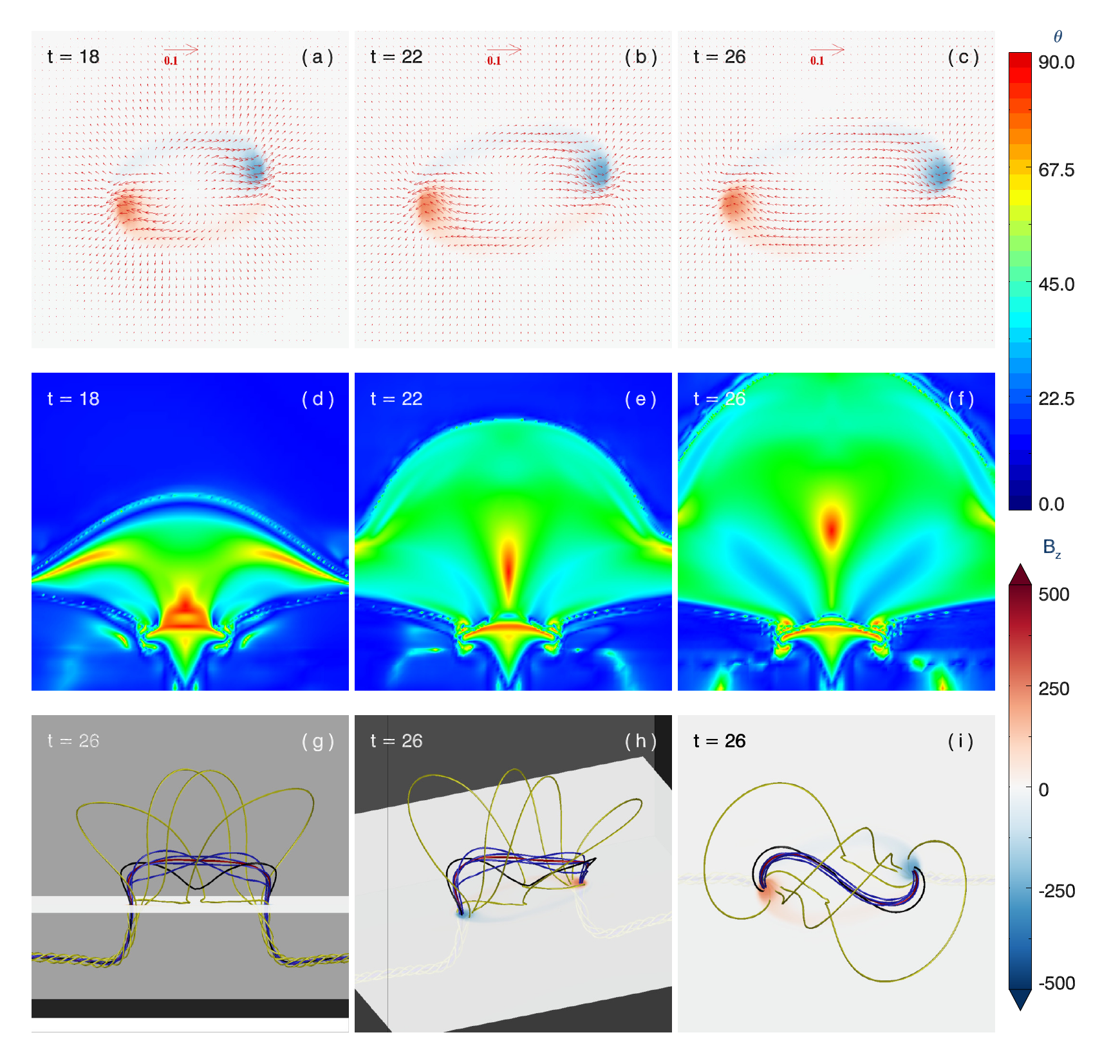}
  \caption{(a-c) The evolution of the $z$-component of the magnetic field ($B_z$, color) and the tangent velocity (arrows) on the surface. (d-f) show the distribution of the shear angle $\theta$ of the emerging magnetic field on the central vertical plane ($x = 0$). (g-i)  The yellow and black lines are the same as in Fig.\ref{Fig:Bline}, and the blue and red lines are the coronal MFR at the position of  minimal $\theta$  on the central vertical plane.}
  \label{Fig:theta}
\end{figure*}

\Fig~\ref{Fig:Bline}(a-l) shows three perspective views of 3D structure and evolution of the magnetic flux tube during the emergence. The black line in these panels, which represents the axis of the initial flux tube, is obtained by tracing the O-point ($B_{\theta}$ minimum) on a vertical cross section of the flux tube at different times. Here the cross section is selected as being the right $x$ boundary, since at its two ends the flux tube evolves much more slowly and is more regular than its middle part that emerges into the atmosphere. The yellow lines are the field lines through four points evenly distributed on this cross section with a small radial distance of $0.02 L_s$ from the O-point. Note that the two ends of the flux tube also expand and evolve (but very slightly) during the emergence of its central portion. Therefore, these field lines are not exactly the same set of field lines in the different panels (or times). Nevertheless, they are a good approximation of the same set of field lines and can reflect the topology and its evolution of the magnetic field. The horizontal slice in each panel represents the solar surface and the color indicates the $z$-component of the magnetic field ($B_z$).

The first column of \Fig~\ref{Fig:Bline} is the snapshot at $t=10$, when the middle of the flux tube has undergone a bulge into an $\Omega$-shape, and then at $t=15$ (the second column in \Fig~\ref{Fig:Bline}), the front of the $\Omega$-shaped flux  has emerged into the atmosphere with a simple arcade configuration, and the central axis magnetic line (black field line) is in a weakly forward S-shape. With time goes on, the emerging flux rapidly expands to the higher corona while the magnetic field structure becomes more complex, and eventually more fluxes emerge forming a strongly reverse S-shaped, i.e., a sigmoid shaped magnetic structure.

\begin{figure}[htbp]
  \center
  \includegraphics[width=8.5 cm]{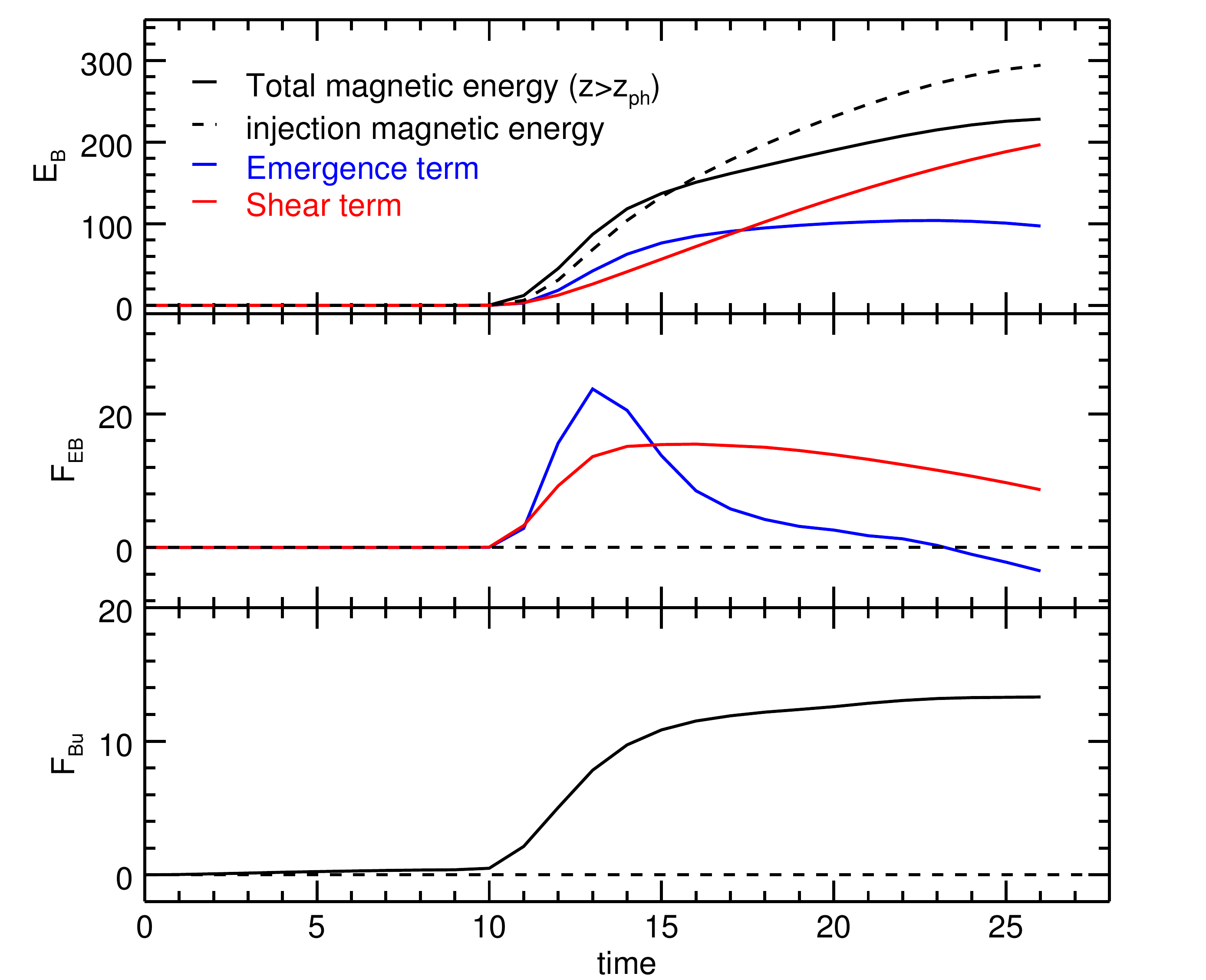}
  \caption{The evolution of magnetic energy ($E_{B}$, top panel), magnetic energy flux ($F_{EB}$, middle panel) and unsigned magnetic flux ($F_{Bu}$, bottom panel). The solid black line in top panel indicates total magnetic energy and the dashed line indicates the injection magnetic energy through the surface, which is the sum of shear term and emergence term of $E_{B}$. The blue and red lines denote the shear term and the emergence term, respectively.}
  \label{Fig:energy}
\end{figure}

\Fig~\ref{Fig:Bline}(m-p) shows the iso-surface of the flux tube with magnetic field strength $B = 0.1B_0$.  At $t=10$, the apex of the flux tube convex part reached the height of the surface. Then at $t=15$, part of the magnetic field has entered the atmosphere in a flattened spherical shape, which indicates that the lateral expansion of the emerging flux is faster than the vertical expansion. With the emergence of flux, the coronal magnetic field also expands wider and higher, eventually forming a ``mushroom" shape.

\subsection{Vortical and shearing motion}

The gradual separation of the two photospheric magnetic polarities as the flux tube emerges can be observed on the horizontal slice (surface) in \Fig~\ref{Fig:Bline}. \Fig~\ref{Fig:theta}(a-c) shows the evolution of the tangent velocity on this slice. These snapshots reveal counterclockwise vortical and shearing motion in each polarity as the flux emerges. It has been suggested that this vortical motion is caused by the difference in the degree of twist $q$ between the subsurface flux tube and the emerged field \citep{Fan2009, Longcope2000AMF}. The expansion and stretch of the emerged flux in corona causes its $q$ to decrease rapidly, and the vortical motion of the two polarities transports the twist of the subsurface flux tube into the atmosphere until the $q$-value equilibrates.

During the evolution of the coronal magnetic field, the combined effect of the vortical motion of the two polarities and the shearing flow distorts the field lines of the emerging flux, turning it from an initial S-shape to an reverse S-shape. The photospheric shearing flow squeezes the bottom of the coronal magnetic field toward its middle, and it has been suggested that magnetic reconnection occurs directly under the sheared field to produce a coronal MFR \citep{Fan2009}. \Fig~\ref{Fig:theta}(d-f) show the distribution of the shear angle $\theta$ (indicating the angle between the magnetic field and the $y-z$ plane) of the emerging magnetic field on the central vertical plane ($x = 0$). We find that the distorted magnetic field gradually separates from the magnetic field that remains below the photosphere, eventually forming a coronal magnetic structure with a sigmoid shaped inner core of MFR. The newly formed coronal MFR at $t=26$ is shown in \Fig~\ref{Fig:theta}(g-i) (blue and red lines). 

The shearing motion of the polarities provides an important way for the magnetic energy to enter the atmosphere through the photosphere, along with the direct upward injection of magnetic field. To quantify the different contributions from these effects, we calculated the total magnetic field energy  above the photospheric surface ($z = 0.39$)  as well as the Poynting flux through the surface for the shear term and the vertical injection term (or emergence term), respectively, using the formula as derived in \citet{ Kusano2002} and \citet{SolarPh2003SoPh}. As can be seen in \Fig~\ref{Fig:energy}, the total magnetic energy above the photosphere increases first quite fast in time from $t=10$ to $15$, in agreement with the fast increase of the unsigned magnetic flux through the photosphere. After then, the total magnetic energy becomes slower and eventually saturates near the end of the simulation (the top panel of \Fig~\ref{Fig:energy}). And the mismatch of the total magnetic energy and the injection magnetic energy means that the contribution by the dissipation and reconnection of magnetic fields is significant at the later phase, accounting for 22.4 $\%$ of the injection magnetic energy. In the mlddle panel of \Fig~\ref{Fig:energy},  the early injection of magnetic energy is contributed mainly by the emergence term, which however decays quickly after $t=13$, and afterwards the shear term dominates. At the later stage of emergence, i.e., when the unsigned magnetic flux has nearly saturated, the emergence term has decreased to a value close to and even below zero at the end of the simulation. This suggests that a small submerge of the magnetic energy occurs. The shear term also decays, but with a much slower rate than that of the emergence term. The net contribution of these two terms eventually stabilizes the total magnetic energy. This is consistent with the simulation of \citet{Magara2003}.

\subsection{Two step emergence}

Our simulation agrees with many previous simulations that the emergence of flux from the convection zone to the corona experiences a two-step process, known as a ``two-step emergence" mode \citep{Matsumoto1993, Magara2001}. The first step is the rise of the flux tube in the convection zone by magnetic buoyancy. During this period, the rising speed of the flux tube initially increases and then decelerates as the flux approaches the surface. The second step is the evolution of the emerging field into the atmosphere. \citet{Toriumi2010} tested the effect of the amount of flux and the initial field strength on the flux emergence in two-dimensional numerical simulations, dividing the results into ``two-step emergence", ``direct emergence" and ``failed emergence". Direct emergence means that the rise of the flux tube is not reduced before breaking through the photospheric surface. Failed emergence is that the flux tube eventually fragments in the convection zone and cannot enter into the atmosphere. The work of \citet{Murray2006}   shows that the twist degree $q$ is also a factor affecting the emergence of the flux tube, with larger values of $q$ favoring for emergence. And \citet{Toriumi2011} point out that the criterion for failure emergence is $q = 0.05$ in 2D simulation.

\begin{figure}[htbp]
  \center
  \includegraphics[width=8cm]{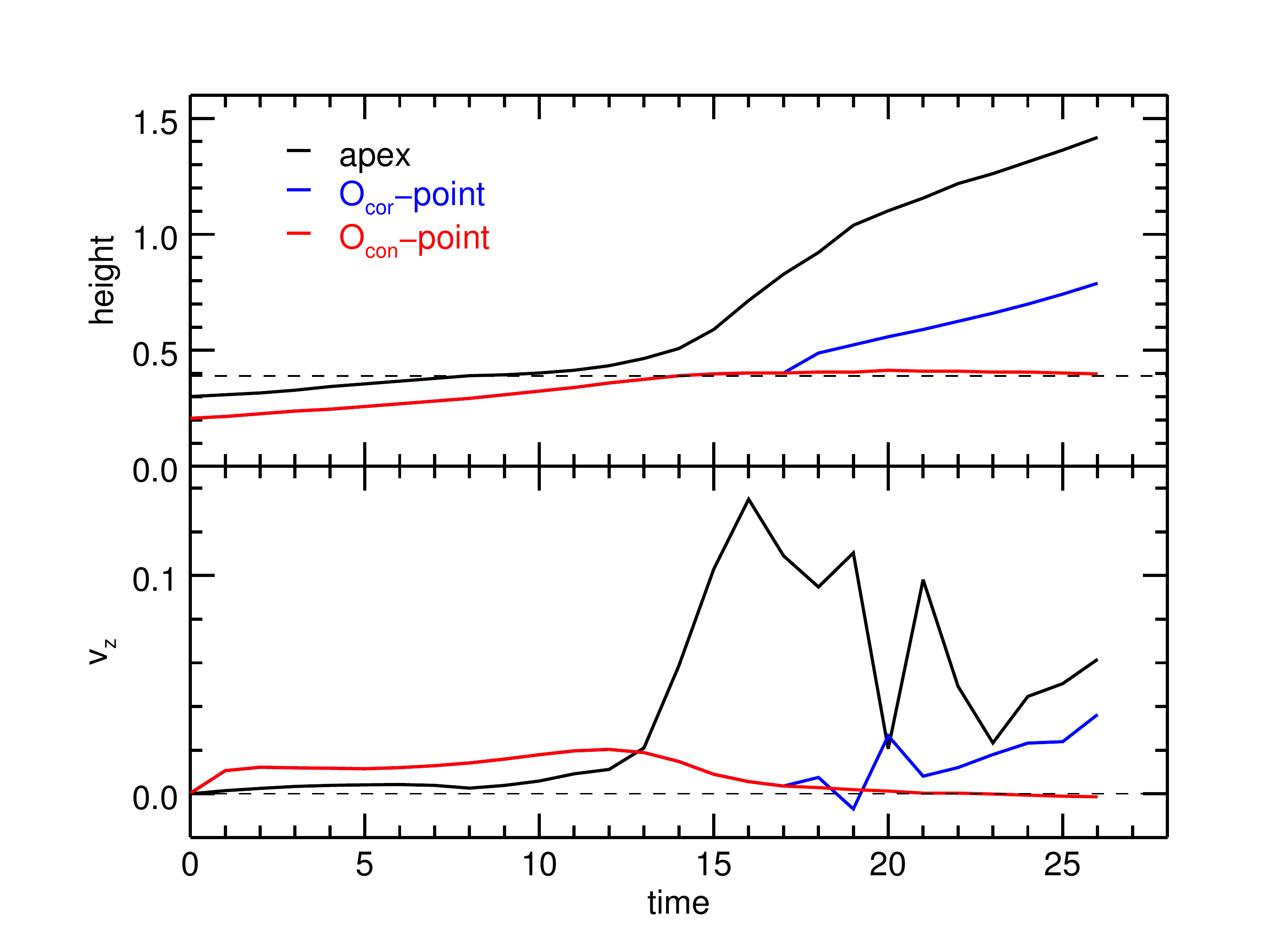}
  \caption{The evolution of the height (top panel) and velocity (bottom panel) of the apex the flux tube and the two O-points in the convection zone and corona on the central vertical plane ($x = 0$). The dash line in the top panel indicates the height of the surface ($z = 0.39$), and the dash line in the bottom panel indicates $v_z = 0$.}
  \label{Fig:trace}
\end{figure}

\Fig~\ref{Fig:trace} shows the evolution of the height (top panel) and velocity (bottom panel) of the apex of the flux tube and the two O-points in the convection zone and corona on the central vertical plane ($x = 0$). Here the apex (black line) is defined to be the highest point where the magnetic field strength $B$ is greater than $0.1B_0$. The evolution of the magnetic flux tube on its middle section is more complicated than that at the boundary, since there are multiple positions of very small $B_{\theta}$ generated during emergence, and thus that the center of the magnetic flux tube on this plane cannot be defined using the same way as in \Fig~\ref{Fig:Bline}. We consider the location with the largest $B_x$ in the minimal $B_{\theta}$ positions on this cross section as the O-point of the flux tube in the convection zone, while the highest position with the minimal $B_{\theta}$ is considered as the O-point of the coronal MFR, denoted by the ${\rm O_{\rm con}}$ and ${\rm O_{\rm cor}}$ points, respectively.

The velocity at the apex of the flux tube (black line) in the bottom panel of \Fig~\ref{Fig:trace} undergoes a process of increase, then decrease, and again increase. The position where the velocity decreases is near the solar surface ($z = 0.39$), thus our simulation belongs to the ``two-step emergence" as defined in \citet{Toriumi2010}. The difference in position between the red and black line in the top panel of \Fig~\ref{Fig:trace} can also reflect the first slow rise, flux pileup near the photosphere, and rapid expansion of the upper part of the flux tube in the corona. \Fig~\ref{Fig:Bline} (m-p) shows that the emerging magnetic field exhibits a significant horizontal expansion, which is one of the key features of the ``two-step emergence". However, in the top panel of  \Fig~\ref{Fig:trace}, the ``pileup" of the apex of the flux tube (black line) near the surface is not obvious, and we consider that it is due to the relatively large $q$ and $B_0$.

In the first step of emergence, the buoyancy of the flux tube is suppressed near the photosphere due to the convective stability of the stratification there, which has a much smaller temperature gradient than that for convection instability \citep{Cheung2007}. Consequently, more and more magnetic fluxes with the frozen plasma that rise from below accumulate near the photosphere, eventually resulting in an unstable configuration in which the heavy plasma (as supported by the magnetic pressure gradient) overlays on the lighter flux tube. Such an unstable configuration is called magnetic buoyancy instability \citep{Matsumoto1993}. \citet{Archontis2004} and \citet{Hood2012b} have given the following critical condition for this instability

\begin{figure}[htbp]
  \center
  \includegraphics[width=8cm]{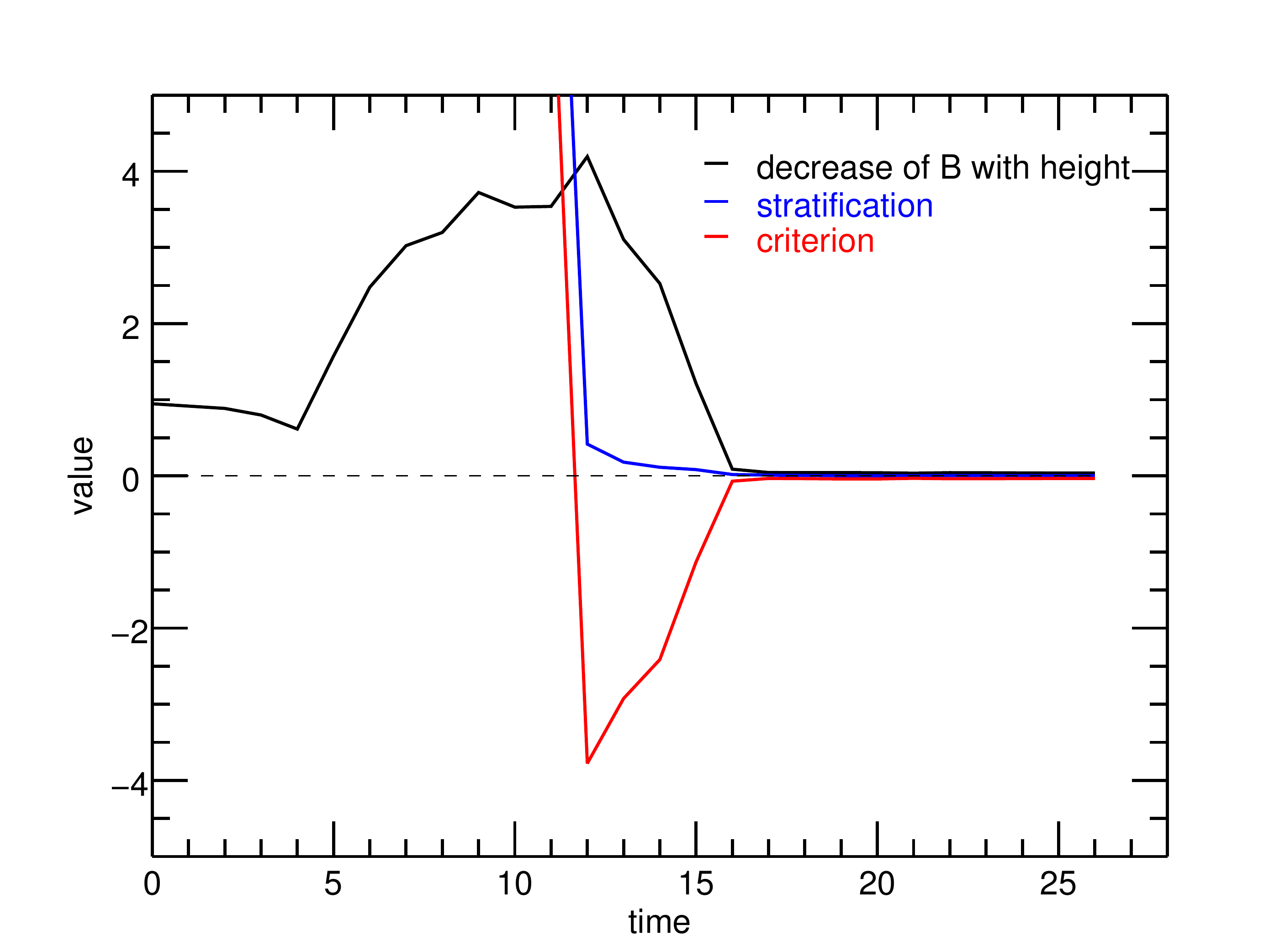}
  \caption{The criterion of magnetic buoyancy instability (red line) for the front of the tube at each moment. The black line describes the variation of the magnetic field strength of the flux tube with height, and the blue line is the stratification effect of the atmosphere. }
  \label{Fig:criterion}
\end{figure}

\begin{figure}[htbp]
  \center
  \includegraphics[width=8.5cm]{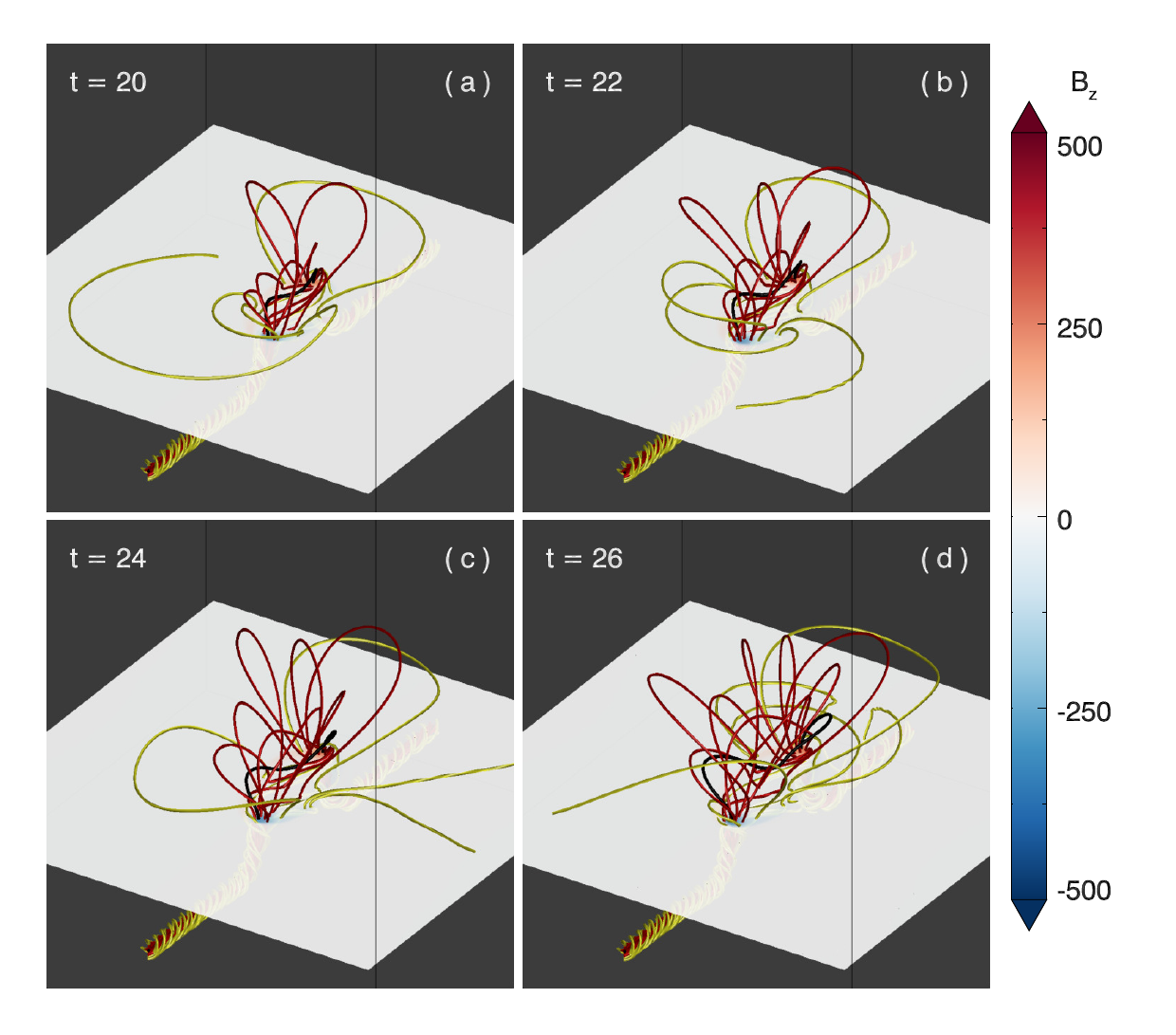}
  \caption{The evolution of the magnetic field lines of flux tube. The transparent horizontal slice represents the solar surface. }
  \label{Fig:centerline}
\end{figure}

	\begin{equation}
      \label{eq:crit}
	-H_p \dfrac{\partial }{\partial z} \left( \log B \right) > -\dfrac{\gamma}{2} \beta \delta + k^2_{\parallel} \left(1+  \dfrac{k^2_z}{k^2_{\perp}}\right),
	\end{equation}
where $H_p$, $z$, $B$, $\gamma$ and $\beta$ denotes the local pressure scale height at the photosphere, the height, the magnetic field strength, the ratio of the specific heats and the ratio of the plasma pressure to the magnetic pressure, respectively. $\delta$ is the superadiabatic index, which is $-0.4$ for a strong stabilization of the atmosphere. $k_{\parallel}$, $ k_{\perp}$, $k_{z}$ are the three components of the local perturbation wave vector. The left side of the equation describes the variation of the magnetic field strength of the flux tube with height, the first term on the right side indicates the stratification effect of the atmosphere, and the second term indicates the effect of the perturbation. This criterion helps us to determine the time of appearance of the flux tube on the surface, since the magnetic flux can only emerge across the surface with the criterion satisfied. We calculated the criterion for the front of the tube at each moment and plotted the result in \Fig~\ref{Fig:criterion}. The equation perturbation term is not shown in the figure since it is a small quantity, that has already included in the criterion (red line). We find that \Eq~(\ref{eq:crit}) is met at $t=12$, indicating that the buoyancy instability is triggered at around this moment, and indeed the magnetic flux first appears above the photosphere between $t=11$ and $12$. It is worth noting that the actual height of the solar surface islifted up by the rising flux tube, thus at $t=11$ the magnetic flux tube exceeds the initial height of the photosphere but is still suppressed by the stability of the stratification.

\subsection{Partial emergence }
\label{sec:pe}
Our simulation also agrees with the existing theory that the magnetic flux tube in the convection zone can only partially emerge into the atmosphere, and the field lines behave on the central vertical plane ($x = 0$) as described in \cite{Leake2013}, i.e., the up-concave part can expand into the corona, while the down-concave part under the original tube axis remains mostly trapped under the surface. To give more details, \Fig~\ref{Fig:centerline} shows the evolution of the field lines traced $17$ points within $0.04 L_s$ near the O-point on the cross section at right $x$ boundary.  These points are the O-point and $4$ points uniformly in each direction along the positive and negative directions of $y$ and $z$, respectively, from O-point. The black line in each panel is obtained by tracing the O-point on right $x$ boundary. The red lines indicate the field lines in the center part of the tube while the yellow lines indicate the outer field lines. 

\begin{figure}[htbp]
  \center
  \includegraphics[width=8cm]{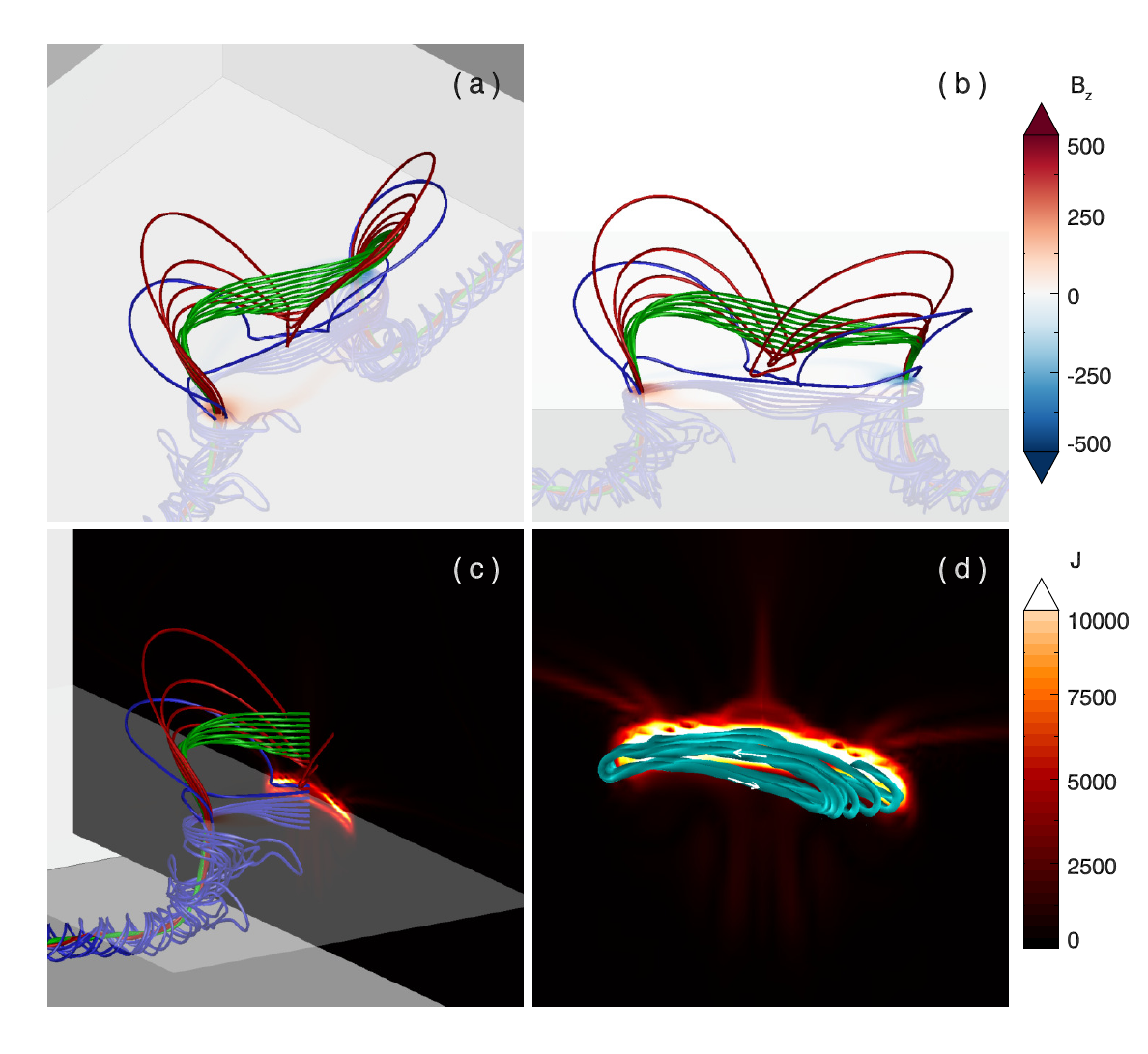}
  \caption{(a-b) Two perspectives of the field lines traced at 20 points uniformly distributed in the height range 0.3$L_s$ to 0.6$L_s$ on the central vertical line at $t = 26$. (c) Add the slice $y=0$. The color indicates the density of current ($J$). (d) The current sheets on the central vertical plane ($x = 0$). The cyan line is the streamline of electric current of the current sheet. }
  \label{Fig:straline}
\end{figure}

\begin{figure}[htbp]
  \center
  \includegraphics[width=8cm]{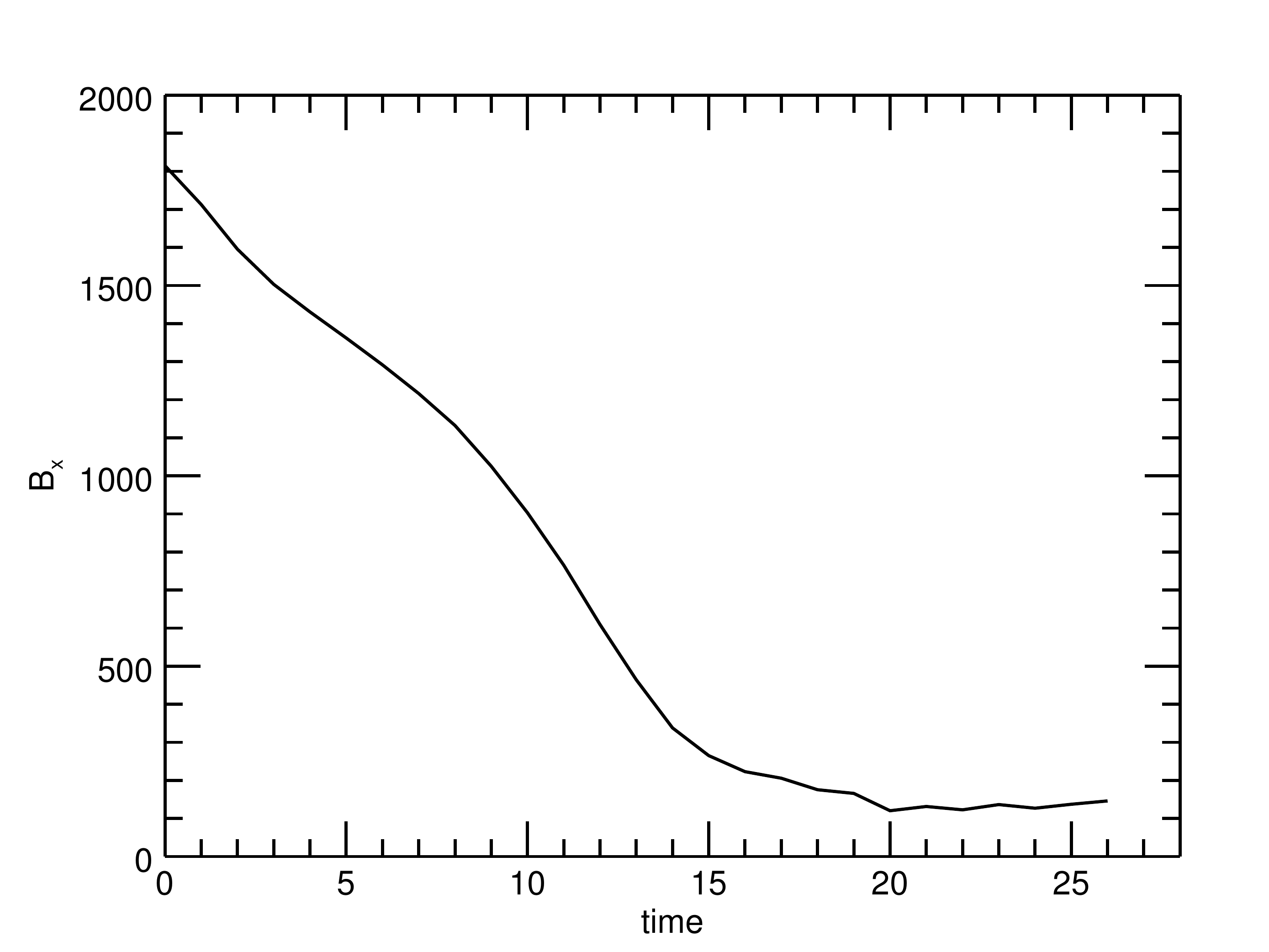}
  \caption{The evolution of $B_x$ at the ${\rm O_{\rm con}}$-point. }
  \label{Fig:Bxtime}
\end{figure}

Similar to the simulation of \citet{Magara2004}, the outer field lines of the emerging flux tube spread out in a wide fan after breaking through the surface, and some filed lines even have a downward trend, such as the yellow field lines at $t = 24$ (\Fig~\ref{Fig:centerline}(c)). The lateral expansion of the inner field lines is restrained by the adjacent twisted field lines, which makes inner field lines tend to expand vertically. With time, the internal field of flux rises to higher corona to form MFR, and remains well connected to the convection zone flux tube.

\begin{figure*}[htbp]
  \center
  \includegraphics[width=18cm]{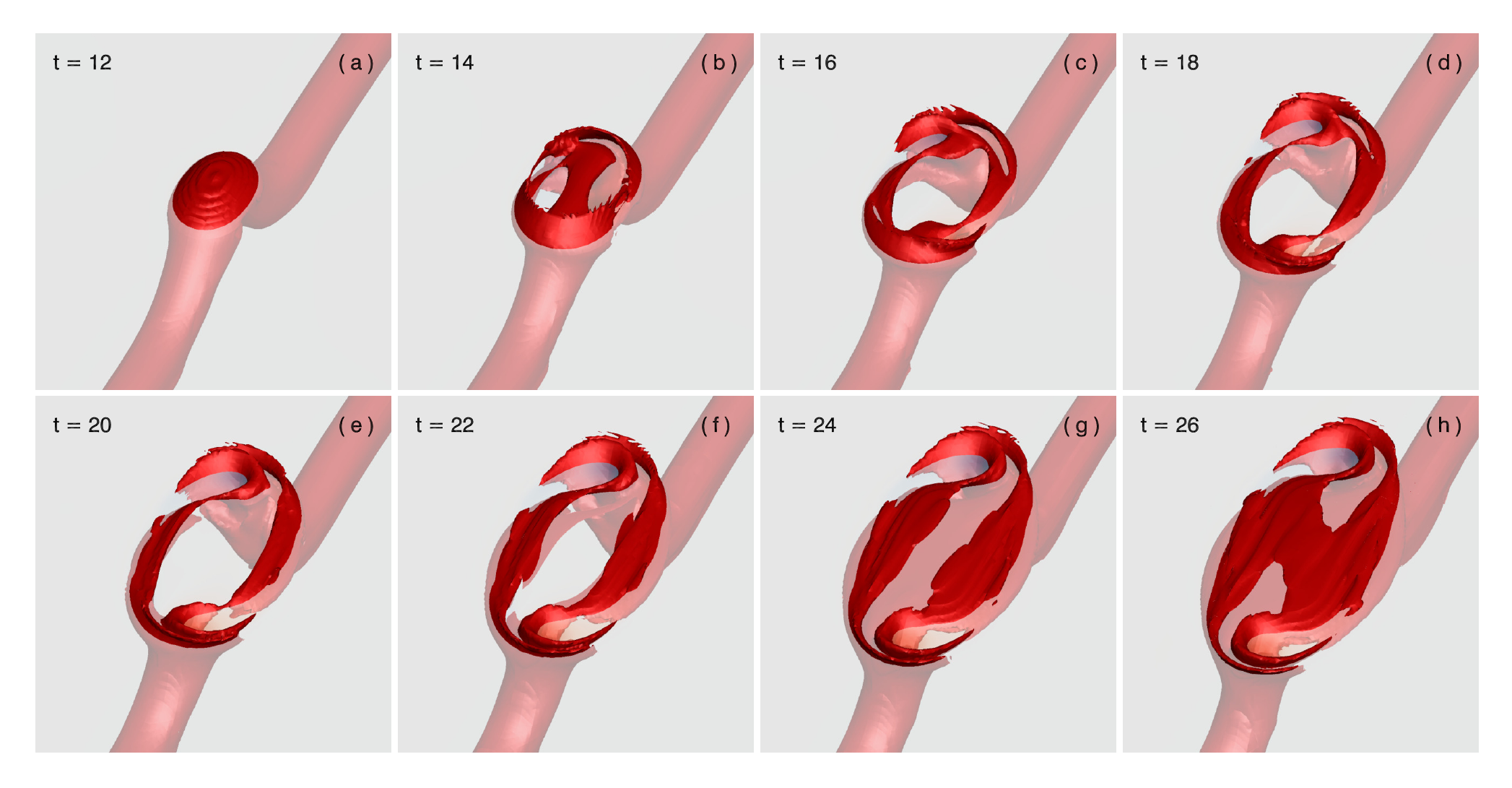}
  \caption{The evolution of the current sheet iso-surface $J=8000$, and the transparent horizontal slice represents the solar surface. }
  \label{Fig:J}
\end{figure*}

\begin{figure}[htbp]
  \center
  \includegraphics[width= 8.5cm]{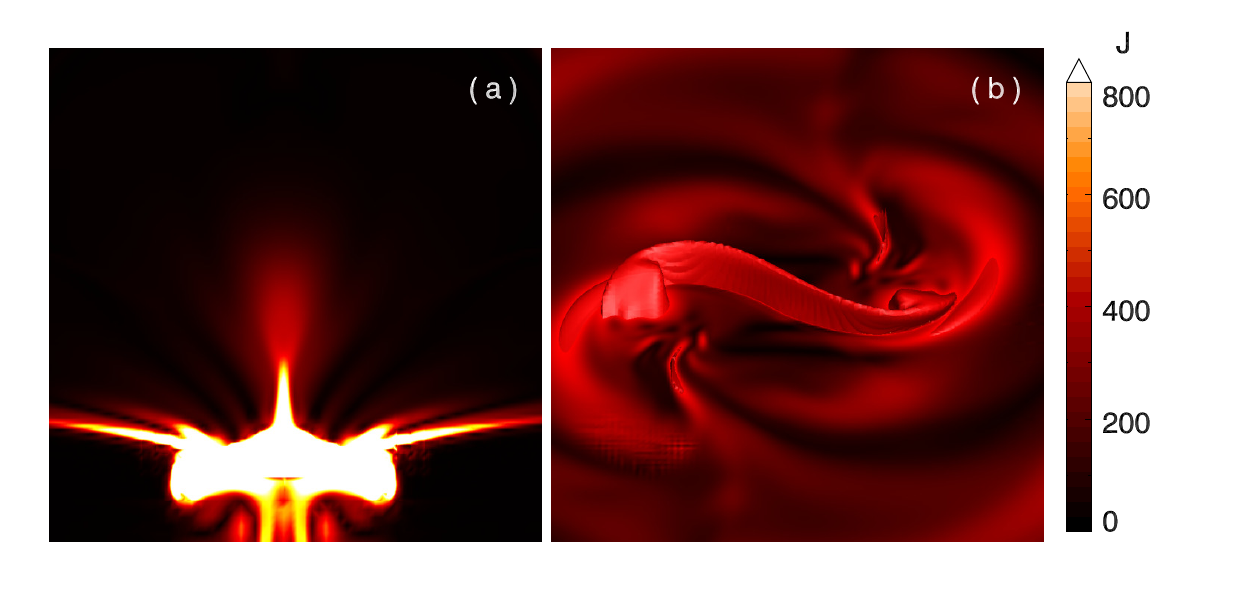}
  \caption{(a) The distribution of the current sheets on the central vertical plane ($x = 0$). (b) The iso-surface of the current sheet $J=300$. }
  \label{Fig:J0}
\end{figure}
\subsection{Current sheet }
\label{sec:cs}

\Fig~\ref{Fig:straline}(a-c) shows the field lines traced at $20$ points that uniformly distributed in the height range $0.3L_s$ to $0.6L_s$ on the central vertical line $(x, y) = (0, 0)$ at $t = 26$. The green lines are the field lines above the black line (same as in \Fig~\ref{Fig:centerline}), which have reverse S-shaped in the corona, with the middle part concave downward. The red lines indicate the magnetic field between the surface and the black line, and the blue lines are the field lines that do not fully emerge.
\citet{Archontis2004} pointed out that the plasma moves along the magnetic line of motion towards the lower part of the field line, and the heavy plasma gathered in the lower part increases the plasma $\beta$, pulls the field lines toward the surface (becoming the structure of the red line in the \Fig~\ref{Fig:centerline}) and reduces the magnetic field gradient, which can cause the convective stability of the stratification to increase. That restrains further emergence of the flux tube in the middle region between the two polarities, resulting in the subsurface field lines in this region not breaking through the photospheric surface.

Although the further emergence of the flux tube is suppressed, the magnetic field can still enter the atmosphere through the motion of the coronal MFR footpoints, which creates the structure of the field lines like the top two of the blue lines in \Fig~\ref{Fig:straline}. The blue and red field lines constitute an X-shaped magnetic field structure in the middle of the two polarity concentration regions, which induces a transverse current sheet. This current sheet is in contact with the current sheet of the subsurface original magnetic flux tube to form a ring current sheet (\Fig~\ref{Fig:straline}(c)), and \Fig~\ref{Fig:straline}(d) shows the streamline of electric current (cyan line) of the ring current sheet. We found that its induced magnetic field is in the same direction as the original flux tube, i.e., it will reduce the tendency of the original magnetic field decay.

\Fig~\ref{Fig:Bxtime} shows the evolution of $B_x$ with time at O-point of the convection zone flux tube, where $B_x$ is hardly decreasing after $t = 20$, which is significantly different from the rapid decrease in the earlier period. This process implies that in the absence of a covered coronal field, the axial direct current is enhanced during the flux emergence and no return current is observed. \citep[for more details on the study of current sheets in simulations of the covering field see][]{Torok2014}.
\Fig~\ref{Fig:J} shows the evolution of the current sheet iso-surface $J=8000$, and the transparent horizontal slice represents the solar surface. We found that in the second step of flux emergence, the evolution of the current sheet is divided into two stages. The first stage is before $t = 20$, when the rapid emergence of partial fluxes causes the subsurface current density to decrease. The second stage is when the current sheet starts to reform and the subsurface current sheet reforms more rapidly. We believe that the formation of the subsurface current sheet is due to the suppression of the heavy plasma in the middle of the two polarity concentration regions, which causes the convection zone magnetic field to accumulate heavily under the surface. The current sheet above the surface is due to the X-shaped magnetic field structure in \Fig~\ref{Fig:straline}. These two current sheets eventually form a cavity configuration.

In addition, the red field lines in \Fig~\ref{Fig:straline} are pulled by the shear flow along the polarity reversal line and the heavy plasma, causing the sides of the coronal magnetic field to squeeze toward the middle, and forming a vertical current sheet, as shown in \Fig~\ref{Fig:J0}. In the real case, the resistivity in the corona is extremely low and reconnection is difficult to occur, which leads to a close reverse magnetic field on both sides and forms a thinner and thinner current sheet that accumulates more and more energy. Once reconnection occurs, a rapid eruption might be produced in the same way as shown in \citet{Jiang2021} that a continuously sheared bipolar arcade can initiate an eruption by tether-cutting reconnection.

\section{Summary}
\label{sec:sum}

In this paper we have implemented the FES using the AMR--CESE--MHD code and has achieved consistent results with many previous FESs of similar configuration but using different numerical codes. The AMR--CESE--MHD method has its uniqueness that it is much simpler in algorithm than traditional numerical MHD solvers but can achieve higher resolution. Further aided with the AMR, it can handle well the drastic variations of many orders of magnitudes in both spatial and time scales in the computational domain that includes the convection zone and the different layers of the solar atmosphere. The computational cost is moderate with around 31 hours on 480 CPUs of 3GHz. 

The simulation follows the whole process of the rising into the corona of a twisted flux tube that is initially placed in the convection zone. As driven by the magnetic buoyancy, the center part of the tube rises until it reaches the photospheric layer. At this position, the reduced gradient of the background temperature produces a stratification stabilization effect, which inhibits the further rise of the flux tube and the magnetic flux starts to pile up near the surface. When the accumulated magnetic field is sufficient to trigger the magnetic buoyancy instability, the upper part of the flux tube begins to emerge into the solar atmosphere and expands rapidly. The emerged magnetic field also suppresses the emergence of the following magnetic field, making only a portion of the original flux tube emerge.

During the evolution of the emerging magnetic field in the corona, vortical and shearing motions of the magnetic polarities on the photosphere play an important role in transporting the magnetic energy and non-potentiality into the atmosphere. To store this energy, the coronal magnetic field has also been reshaped to a sigmoid configuration (containing a weakly twisted rope) from the simple arcade at the early time of the emergence. Due to the strong lateral expansion of the coronal field, the entire 3D profile of the coronal field resembles the shape of a ``mushroom''. 

In addition, we also analyze the formation of the current sheet. The shear flow of the photospheric layer squeezes the sides of the coronal magnetic field toward the middle, and the reversed magnetic field (as seen on the central cross section) gets closer and closer, leading to the formation of a vertical current sheet. We also found that below this vertical current sheet, the horizontal current sheet on the surface forms a cavity structure with the current sheet in the convection zone, and the presence of the toroidal current increases the magnetic field in the convection zone, which may lead to the re-emergence of the magnetic field \citep{Syntelis2017}.

The present work developed a framework for numerical experiments of magnetic flux emergence and its role in producing solar eruptions, which will be the focus of our future works. For example, with an ultra-high accuracy MHD simulation, \citet{Jiang2021}  established a fundamental mechanism behind solar eruption initiation: a bipolar field driven by slow shearing motion on the photosphere can form an internal current sheet in a quasi-static way, which is followed by fast magnetic reconnection (in the current sheet) that triggers and drives the eruption. However, their model domain includes only the corona by assuming the lower layers of atmosphere below the coronal base (i.e., the photosphere and chromosphere) as a line-tied boundary surface, and the surface driving velocity is also specified in an ad-hoc way. This inspires us to perform higher resolution FES to investigate whether the same mechanism can also operate to produce eruptions during the evolution of the emerging flux in the corona, with the shearing motion at the photosphere generated in a more self-consistently way. In another study, \citet{Bian2022} showed that by the continuous shearing of the same PIL, the fundamental mechanism can effectively produce homologous CMEs by recurring formation and disruption of the internal current sheet. Such homologous eruptions will be also investigated with longer-term FESs to verify whether a second emergence will occur after the first emergence of the same flux tube. And the FESs have other important applications in studying the solar eruptions, in particular, to explore what are the key parameters that can be used to predict eruptions. One of the such applications has been shown by \citet{Pariat2017} who used the FESs by \citet{Leake2013} and found that the ratio of the magnetic helicity of the current-carrying magnetic field to the total relative helicity can potentially used for eruption prediction. This merits further studies using our FES model.


\acknowledgments This work is jointly supported by the National Natural Science Foundation of China (NSFC 42174200 and 41731067), the Fundamental Research Funds for the Central Universities (HIT.OCEF.2021033), and the Shenzhen Science and Technology Program (RCJC20210609104422048 and JCYJ20190806142609035). The computational work was carried out on TianHe-1(A), National Supercomputer Center in Tianjin, China.

\bibliographystyle{apj}
\bibliography{all}

\end{document}